\documentclass[conference]{IEEEtran}
\IEEEoverridecommandlockouts
\usepackage{cite}
\usepackage{amsmath,amssymb,amsfonts}
\newtheorem{definition}{Definition}
\usepackage{algorithmic}
\usepackage{graphicx}
\usepackage{textcomp}
\usepackage{xcolor}
\usepackage{seqsplit}
\usepackage{tikz}
\usepackage{tabularx}
\usepackage{array}
\usepackage{listings}
\usepackage{adjustbox}
\usepackage{multirow}
\usepackage{pgfplots}
\pgfplotsset{width=10cm,compat=1.16}
\newcolumntype{M}[1]{>{\centering\arraybackslash}m{#1}}
\usetikzlibrary{shapes.geometric,shapes.symbols,fit,positioning,shadows, arrows.meta, backgrounds}

\def\BibTeX{{\rm B\kern-.05em{\sc i\kern-.025em b}\kern-.08em
		T\kern-.1667em\lower.7ex\hbox{E}\kern-.125emX}}
\begin{document}
	\makeatletter
	\newcommand{\linebreakand}{%
	\end{@IEEEauthorhalign}\vspace*{-0.3cm}
	\hfill\mbox{}\par
	\mbox{}\hfill\begin{@IEEEauthorhalign}
	}
	\makeatother
	\title{Exploring Weighted Property Approaches for RDF Graph Similarity Measure
		
		\vspace*{-0.3cm}
	}
	\author{
			\IEEEauthorblockN{Ngoc Luyen Le\IEEEauthorrefmark{1}\IEEEauthorrefmark{2}\\\textit{ngoc-luyen.le@hds.utc.fr}}
			\and
			\IEEEauthorblockN{Marie-Hélène Abel\IEEEauthorrefmark{1}\\\textit{marie-helene.abel@hds.utc.fr}}
			\and
			\IEEEauthorblockN{Philippe Gouspillou\IEEEauthorrefmark{2} \\\textit{p.gouspillou@vivocaz.fr} }
			\linebreakand
			\IEEEauthorblockA{
				\IEEEauthorrefmark{1}Université de technologie de Compiègne, CNRS, Heudiasyc (Heuristics and Diagnosis of Complex Systems),\\ CS 60319 - 60203 Compiègne Cedex, France.
				\\
				\IEEEauthorrefmark{2}Vivocaz, 8 B Rue de la Gare, 02200, Mercin-et-Vaux, France.
			}
			\vspace*{-1.1cm}
	}

	
	\maketitle

	\begin{abstract}
		
		Measuring similarity between RDF graphs is essential for various applications, including knowledge discovery, semantic web analysis, and recommender systems. However, traditional similarity measures often treat all properties equally, potentially overlooking the varying importance of different properties in different contexts. Consequently, exploring weighted property approaches for RDF graph similarity measure presents an intriguing avenue for investigation. Therefore, in this paper, we propose a weighted property approach for RDF graph similarity measure to address this limitation. Our approach incorporates the relative importance of properties into the similarity calculation, enabling a more nuanced and context-aware measures of similarity. We evaluate our approach through a comprehensive experimental study on an RDF graph dataset in the vehicle domain. Our results demonstrate that the proposed approach achieves promising accuracy and effectively reflects the perceived similarity between RDF graphs.
		
	\end{abstract}
	
	
	\section{Introduction}
	
	The analysis and representation of data in the form of RDF (Resource Description Framework) graphs have become increasingly prevalent in various domains, ranging from semantic web applications to knowledge representation and graph databases \cite{bizer2011linked, hernandez2015reifying}. RDF graphs serve as powerful models for capturing complex relationships and structured information, making them essential tools for representing knowledge in a structured and interconnected manner \cite{auer2007dbpedia}. These graph-based structures have proven instrumental in semantic web technologies, facilitating data integration and providing a foundation for reasoning and inference \cite{world2023rdf}.
	
	One of the fundamental tasks in RDF data management and knowledge integration is the assessment and comparison of RDF graphs for their degree of similarity. Evaluating the similarity between RDF graphs holds essential significance across a spectrum of applications, encompassing information retrieval, knowledge discovery, semantic web analysis, and recommender systems \cite{harispe2015semantic, petrova2017entity, luyenln_improving}. This process serves as a fundamental means to identify and link related entities, discover patterns in data, and enable more accurate and context-aware decision-making processes.
	
	Contemporary approaches to RDF graph similarity predominantly focus on structural properties, encompassing the topology and connectivity of nodes and edges, and often treat all properties equally \cite{maillot2018measuring, tartari2018wisp, harispe2013semantic}. However, these methodologies can fall short in fully capturing the intricate semantic subtleties inherent in the data. Real-world RDF graphs often include a myriad of properties intricately linked to nodes and edges, which significantly influence data interpretation and relevance. It is important to note that within RDF graphs, the importance of individual properties can vary in different contexts. In specific scenarios, certain properties may play a more crucial role than others, depending on the purpose of the RDF graph comparison.
	To address this limitation, we investigate to explore weighted property approaches for RDF graph similarity measure. Property weights enable the expression of the relative importance or significance of distinct properties within the RDF graph, facilitating more nuanced and context-aware measures of similarity.
	
	In the context of this study, we propose an exploration of weighted property approaches for measuring RDF graph similarity, delving into the evolving landscape of RDF graph similarity measure. Our study incorporates the relative importance of properties into the similarity calculation, enabling a more nuanced and context-aware measures of similarity. As results, we evaluate our approach through a comprehensive experimental study on an RDF graph dataset in the vehicle domain. By doing so, we aim to meet the demand for more accurate and context-sensitive RDF graph comparisons, particularly in scenarios where property information plays a important role in data filtering and knowledge discovery

	The remainder of this paper is organized as follows: In the following section, we introduce literature on RDF knowledge graphs and the approach for measuring similarity. Section \ref{section_relatedwork} presents our main works to the formulations and our proposed method for exploring weighted properties in measuring similarity on RDF graphs. In Section \ref{proposition}, we experiment with our proposed approach based on an RDF knowledge graph in the vehicle purchase/sale domain. Finally, we conclude the paper with some ideas for future work in the last section.
	
	\section{Related Work}\label{section_relatedwork}
	In this section, we will explore RDF data graphs and examine various methods for measuring RDF graph similarity. Our aim is to synthesize the advantages and drawbacks of these similarity measurements.
	
	
	Fundamentally, RDF data, often based on ontological model, serves as a framework for structuring information. RDF facts are defined by triples, each consisting of three components: a subject, a predicate, and an object. Essentially, a triple denoted as $\langle subject, predicate, object \rangle$ signifies that a given subject is associated with a specific value for a particular property\cite{abiteboul_2011, luyen2016development}. RDF graph data inherently possesses the ability to make inferences and reason about additional knowledge, particularly when associated with ontologies. Semantic similarity within RDF graph data refers to the proximity of two terms\footnote{A term can represent a concept, subject, predicate, object, or even a set of triples} within the structure of the RDF graph data itself. The measure of distance between two terms is represented numerically as a vector, indicating their closeness to each other \cite{lee2008comparison}. This capacity facilitates the efficient utilization of RDF graph data for retrieving related elements or discerning associations between terms.
	
	Several works have explored the concept of weighted properties for enhancing RDF graph searching. For instance, the author in \cite{liu2012r2db} suggests a special way to handle and search weighted RDF graphs, where different properties in the data are given different levels of importance. This system not only helps in searching these graphs but also allows you to see the information you find, making it easier to understand complicated relationships in the data. Another work by the authors in \cite{song2019construct} introduces a new SPARQL extension that uses the CONSTRUCT clause to support analytics on weighted RDF graphs. With this approach, users can perform complex tasks like calculating PageRank for influence analysis, taking into account the different weights assigned to properties in the graph. These approaches recognize the limitations of traditional methods that treat all properties equally. However, they differ in how they include weights and the aspects they focus on. Some papers might delve deep into the theory behind weighted properties, while others might suggest specific ways to assign weights or explore practical uses.

	Evaluating the similarity between RDF graphs has been an active area of research for several years, with various approaches proposed in the literature. Depending on the structural characteristics of the application context and the chosen knowledge representation model, researchers and practitioners have introduced various similarity measures\cite{meym2016semantic,sanchez2012ontology}. These approaches can be broadly classified into four primary categories: (i) path-based, (ii) feature-based, (iii) information content-based, and (iv) hybrid approaches. (i) Path-based approaches conceptualize RDF graphs as directed graphs, with nodes and links representing entities interconnected through hypernym and homonym relationships. This approach implements a hierarchical organization and assesses semantic similarities by evaluating semantic relationships, with the distance and nature of the paths between entities \cite{meym2016semantic}. Such approaches are commended for their straightforwardness and reduced computational demands, necessitating only essential information regarding each entity, thus enabling rapid and effective similarity evaluations\cite{li2021ontology}. However, their major limitation is their dependence on the quality of the graph's structure, specifically the completeness, homogeneity, coverage, and granularity of defined relationships. (ii) Feature-based approaches focus on the features or properties of the entities within the RDF graphs. It involves comparing these features to determine how similar or different the entities are. This approach is often employed when the properties of the entities carry significant information about their nature or status\cite{meym2016semantic}. Similarity evaluation can be conducted using various coefficients for property sets, including the Dice's coefficient \cite{dice1945} and Jaccard index \cite{jaccard1901etude}. This approach is beneficial as it assesses both the similarities and differences in the compared property sets, thus enabling the extraction of a wider range of semantic knowledge than is possible with the path-based approach. However, a key challenge in this approach is the need to carefully balance each property's contribution. (iii) Information content-based approaches determine similarity by calculating the information content of entities and associating probabilities with their occurrence\cite{sanchez2012ontology}. This method is particularly important when the value of the information each entity carries plays a significant role in assessing similarity. In this context, entities that appear less frequently in an RDF graph are deemed more informative than those that are more common. However, a limitation of this approach is the need for large RDF graphs that possess a detailed taxonomic structure, which is essential for accurately differentiating between various entities. (iv) Hybrid approaches combine elements of the above approaches. These strategies might mix path-based and feature-based methods, or any other combination, to leverage the strengths of multiple approaches \cite{gao2013ontology}. This is particularly useful in complex scenarios where a single type of approach might not be sufficient to capture the nuances of RDF graph similarity.
	
	In the feature-based approach, the significance assigned to each property varies, reflecting the distinct roles these properties play. This variation is crucial as it recognizes that not all properties contribute equally to the overall assessment of similarity or differentiation within the RDF graph, depending on the context for measuring their similarity. Understanding and appropriately weighting these differences is key to the effectiveness of this approach, as it directly impacts the accuracy and relevance of the similarity measures obtained. In the next section, we will explore our proposition for computing similarity between RDF graphs.
	
	\section{Our proposition}\label{proposition}
	We will delve into the formulations surrounding RDF graphs and our proposed method for measuring the similarity between these graphs, with a special emphasis on exploring weighted properties.
	
	\subsection{Formulations}
	We explore the foundational concepts of RDF, beginning with an examination of the RDF triple and the RDF graph. Initially, we will clarify the nature of an RDF triple and investigate how its three components—the subject, predicate, and object—contribute to forming significant statements about various entities. Subsequently, we will examine how these RDF triples interconnect, creating intricate networks of information.
	\begin{definition}
		An RDF triple is three-tuple of the form $t=\langle s, p, o \rangle$, where $s$ is the subject, denoting the entity or resource being described, $p$ is the predicate, representing the property or characteristic of the subject, and $o$ is the object, expressing the value or state of that property.
	\end{definition}
	
	An RDF triple forms a statement or assertion about a resource, encapsulating a singular unit of information or fact. For instance, consider the RDF triple $\langle$``$Tesla$ $Model$ $S$'', ``$Made$ $by$'', ``$Tesla$ $Motors$''$\rangle$. In this structure, the subject is ``$Tesla$ $Model$ $S$'', which is the entity being described. The predicate ``$Made$ $by$'' acts as the relational link, indicating the type of relationship that exists between the subject and the object. Finally, the object ``$Tesla$ $Motors$'' completes the statement, providing the specific detail that the $Tesla$ $Model$ $S$ is manufactured by $Tesla$ $Motors$. This triple-based structure can be used for representing complex networks of information. By linking various triples, a comprehensive and interconnected data is formed. By that way, a collection of RDF triples inherently carries semantic meaning, with each component — subject, predicate, and object — playing a vital role in defining the conveyed information. This emphasis on triples facilitates the representation of complex, semantically rich data structures in RDF graphs, wherein the interaction of subjects, predicates, and objects forms a network of meaningfully interconnected information \cite{sagi2022design}.
	
	\begin{figure}[h!]
		\begin{center}
			\includegraphics[width=0.45\textwidth]{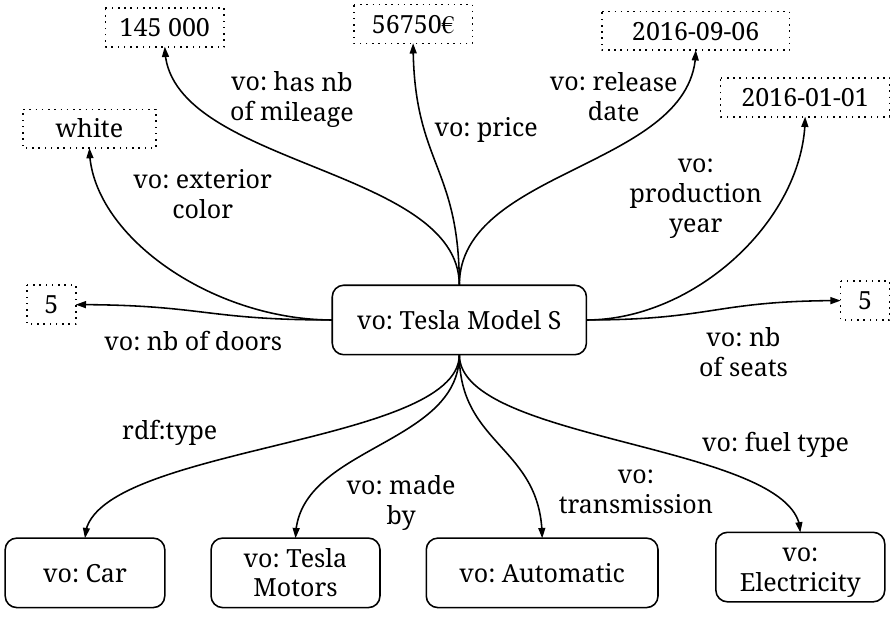}
		\end{center}
		\caption{A snapshot of an RDF graph represents information about a vehicle model (\textit{vo} note for Vehicle Ontology)}
		\label{fig01}
	\end{figure}
	
	\begin{definition}
		Given that $t=\langle s, p, o \rangle$ is an RDF triple, an RDF graph $G$ is defined as a set of such RDF triples $t$, i.e., $G=\{t_1, t_2, ...,t_n\}$, where the interconnections between different triples, through shared subjects or objects, represent the relationships and structures within the graph.
	\end{definition}
	
	This set-based representation of an RDF graph closely mirrors the foundational concepts of graph theory in mathematics. In traditional graph theory, a graph is typically viewed as a collection of vertices (nodes) connected by edges (links). Similarly, in the context of RDF graphs, the vertices can be equated to subjects and objects, while the edges correspond to the predicates that establish a connection between these vertices. For example, Figure \ref{fig01} illustrates a data graph that represents information about the ``\textit{Tesla Model S}'', constructed based on a set of RDF triples. However, a notable distinction in RDF graphs is the emphasis placed on the triples themselves, rather than on the vertices and edges in isolation. This representation enables the flexible modeling of real-world entities and their relationships in a way that is both structured and adaptable. Consequently, RDF graphs are particularly adept at handling diverse and dynamic data sets, making them a powerful tool for data integration, knowledge representation, and information retrieval.
	
	In the field of data analysis and knowledge representation, similarity measures are essential for identifying and comparing entities based on their shared characteristics. RDF graphs, a widely adopted formalism for representing knowledge, have become a powerful tool for modeling complex relationships between entities. The effective comparison of RDF graphs necessitates the use of similarity measures. In the following section, we will explore our method for measuring similarity, focusing on the utilization of weighted properties.
	
	\subsection{Weighted Properties}\label{sec_wp}

	Traditional similarity measures for RDF graphs often assign equal weight to all properties, which may not always reflect the true importance of different properties in determining similarity. To address this limitation, the concept of weighted properties is employed, involving the assignment of varying levels of importance or weights to the properties (or predicates) within the RDF triples. This method acknowledges that properties within RDF graphs do not contribute equally to the semantic meaning or relevance of the data. A weighted property, therefore, is a numerical value assigned to a specific property or relationship between entities in the graphs, indicating the significance of that property in determining the overall similarity. 
	
	\begin{definition}
		A weighted property is defined as a pair ($p$, $w$), where $p$ is the property or relationship between entities, $w$ denotes the weight assigned to the property, representing its importance in similarity measure.
	\end{definition}
	
	By utilizing weighted properties, similarity measures can more accurately reflect the different levels of importance of various properties when determining the overall similarity between RDF graphs. This approach results in similarity assessments that are both nuanced and sensitive to the specific context. Consider an example where weighted properties are used in conjunction with an RDF graph to recommend vehicle models that align with a user's preferences. Suppose the user is seeking vehicles with a $budget$ $of$ $less$ $than$ $\$10,000$, $white$ $color$, $automatic$ $transmission$, and $5$ $seats$. In this case, each vehicle model can be represented as an RDF graph, or a set of RDF triples $G$ as follows:
	
	\begin{tabular}{l}
		\{($\langle$ $Tesla$ $Model$ $S$, $price$, $56750€$$\rangle$, $0.2$)	\\
		($\langle$ $Tesla$ $Model$ $S$, $exterior$ $color$, $white$$\rangle$, $0.2$) \\
		($\langle$ $Tesla$ $Model$ $S$, $transmission$, $automatique$$\rangle$, $0.2$)\\
		($\langle$ $Tesla$ $Model$ $S$, $nb$ $of$ $seats$, $5$$\rangle$, $0.2$)\\
		($\langle$ $Tesla$ $Model$ $S$, $has$ $nb$ $of$ $mileage$, $145 000$$\rangle$, $0.1$)\\
		($\langle$ $Tesla$ $Model$ $S$, $fuel$ $type$, $electricity$$\rangle$, $0.1$)\}\\
	\end{tabular}
	In this example, each vehicle model is represented in an RDF graph with properties corresponding to the predicates of triples: $budget$, $color$, $transmission$, $seating$ $capacity$, etc. The similarity measure employs these weighted properties to score each vehicle based on its alignment with the user's preferences. Vehicles that closely match the higher-weighted preferences will receive a higher score, making them more likely to be recommended to the user. Essentially, this demonstrates the effective use of weighted properties in an RDF graph to provide customized vehicle recommendations tailored to specific user preferences.
	
	By assigning numerical values to each property of entities in the graph to indicate their importance, weighted properties offer several advantages in similarity measure. Firstly, they enhance accuracy by accounting for the different levels of significance of various properties, leading to more precise and dependable similarity assessments. Secondly, they provide context-awareness, as incorporating domain-specific knowledge into weight assignment makes the similarity measures more aligned to the particular context of the application. Thirdly, weighted properties impart adaptability, allowing for the customization of similarity measures to fit various applications and domains, thus ensuring their flexibility. In the following section, we will delve into the methodology of measuring similarity using RDF graphs and weighted properties. This will aim to provide a comprehensive understanding of the methodology, highlighting its systematic approach and underlying principles.
	
	\subsection{Similarity Measure Using Weighted Properties}
	
	In our work, we have developed a hybrid approach for measuring RDF graphs similarity , which incorporates a variety of techniques based on both feature-based and information content-based approach, drawing inspiration from the work of Le and colleagues \cite{luyenln_improving}. Consequently, our process for measuring similarity involves an in-depth comparison of two RDF graphs, breaking them down into both quantitative and qualitative components. For triples that have quantitative objects, we utilize a feature-based similarity approach, which is particularly effective for comparing the objects of the triples by evaluating their numerical values. On the other hand, for triples with textual objects, our focus shifts to comparing the subjects and predicates using an information content-based approach. This method entails analyzing the intrinsic information and meaning within the subjects and predicates, thereby facilitating a more comprehensive understanding of their semantic relationships.
	
	Initially, it's important to note that the objects in these triples are represented by numeric values. Comparing these numbers is straightforward and involves measuring the distance between them. For comparing two distinct objects, the Euclidean distance is utilized. Consequently, the smaller the difference between two objects, the greater their similarity. Let consider two objects, $t_{o1}$ and $t_{o2}$, with their respective vectors being $t_{o1} = \{o_{11}, o_{12}, ..., o_{1d}\}$ and $t_{o2} =\{o_{21}, o_{22}, ..., o_{2d}\}$. The semantic similarity between these objects is then defined as follows: 
	
	\vspace{-0.3cm}
	\begin{equation}
		Sim_1(t_{o1},t_{o2}) = \frac{1}{1 + \sqrt{\sum_{c=0}^{d}(o_{1c} - o_{2c})^2}}
	\end{equation}
	\vspace{-0.3cm}
		
	Subsequently, consider $t_{s1}$ and $t_{s2}$ be two qualitative component in two triples $t_1$ and $t_2$ whose numerical vectors are $M_1 = \{\vec{m_{11}}, \vec{m_{12}}, ..., \vec{m_{1h}}\}$ and $ M_2 = \{\vec{m_{21}}, \vec{m_{22}}, ..., \vec{m_{2l}}\}$, their similarity is defined as follows:
	\begin{equation}
		Sim_2(t_{s1},t_{s2}) = \frac{\sum_{u=1}^{h} \bar{S}(\vec{m_{1u}}, t_{s2}) + \sum_{v=1}^{l} \bar{S}(\vec{m_{2v}}, t_{s1})}{h + l}
	\end{equation}
	In this formula, $\bar{S}(\vec{m}, t_{s})$ signifies the similarity between a word $\vec{m}$ and a qualitative component. The function $\bar{S}(\vec{m}, t_{s})$ is calculated as follows:
	\begin{equation}
		\bar{S}(\vec{m}, t_{s}) = \max\limits_{\vec{m_u} \in M} \bar{S}(\vec{m}, \vec{m_u})
	\end{equation} where $\vec{m_u} \in M=\{\vec{m_1}, \vec{m_2}, ..., \vec{m_h}\}$ represents the word vector of $t_s$ with each word $\vec{m_i}$ expressed as a numerical vector. The Term Frequency-Inverse Document Frequency (TF-IDF) word frequency method is typically employed to create numerical vectors for each word by estimating its occurrence probability within a set of triples. However, this technique exhibits a notable limitation: it does not effectively capture the nuances and the positional context of words within the triple sequence. This limitation stems from its reliance on the frequency of word appearances within a set of triples and across multiple sets. To remedy this, we propose the use of Continuous Bag of Words (CBOW) and Skip-gram models, integrated within the Word2vec pretrained corpus \cite{mikolov2013efficient, abdine2021evaluation}. These models are better equipped to grasp the semantic relationships and the sequential arrangement of words, providing a more contextually rich and semantically informed vector representation.

	Finally, the similarity measure between the two RDF graphs, based on the use of weighted properties,  $G_1=\{\langle t_{11},w_{1}\rangle, \langle t_{12},w_{2}\rangle, ..., \langle t_{1g},w_{g}\rangle\}$ and $G_2=\{\langle t_{21},w_{1}\rangle, \langle t_{22},w_{2}\rangle, ..., \langle t_{2g},w_{g}\rangle\}$ is determined by comparing the similarity of each individual triplet as follows:
	\begin{equation}
		\begin{split}
			&Sim(G_1, G_2)= \\ &\frac{\sum_{i=0}^{q} Sim_{1}(t_{1i},t_{2i})*w_{i} \;+  \sum_{j=0}^{r} Sim_{2}(t_{1j},t_{2j})*w_{j}}{\sum_{k=0}^{g}(w_k)}
		\end{split}
	\end{equation} where $q$ represents the total number of triples containing qualitative objects, and $r$ denotes the total number of triples with quantitative objects in the two RDF graphs $G_1$ and $G_2$.
	
	To accurately measuring the similarity between RDF graphs, it is essential to consider the relative importance of each triplet comparison. This can be achieved by assigning weights $w$ to each triplet comparison, ensuring that the overall similarity score reflects the significance of each triple. This approach is particularly crucial in applications where certain triples may hold greater relevance than others, such as in knowledge discovery or semantic similarity analysis. 
	
	Having established the theoretical framework for measuring RDF graph similarity using weighted properties, the next section will detail our experimental setup, describe the data utilized, and present the results obtained.

	\section{Experiments}\label{experiments}
	In this section, we conduct an empirical test of our approach using an RDF dataset within the vehicle domain \cite{le2021towards}. The dataset chosen for this experiment includes approximately 1000 used vehicle models, with each vehicle model described by a variety of characteristics represented in a set of RDF triples. Our goal in applying this approach is to demonstrate its practicality and efficacy in a domain where diverse vehicle properties are crucial in determining similarity.
	
	Drawing from the collected instances, we have conducted experiments and evaluations on four distinct approaches:
	\begin{itemize}
		\item $PJ$: This approach utilizes the Jaccard index to measure similarities between sets of triplets, as detailed in Fletcher and colleagues' study \cite{fletcher2018comparing}.
		\item $PS$: Proposed by Siying Li and colleagues' study \cite{li2021ontology}, this hybrid method merges strategies based on information content and features. However, it focuses solely on the objects and predicates of the triplets.
		\item $P0$: The approach proposed by \cite{luyenln_improving} involves measuring similarity using word embedding for textual objects.
		\item From $P1$ to $P11$: Our primary proposed approach consists of applying weighted properties to various aspects of vehicle models. In this method, we delve into the significance of different properties in the overall measure of similarity. Accordingly, we assign a greater weight to each specified property as follows: $P1=\{Release\;Year\}$, $P2=\{Mileage\}$, $P3=\{Fuel$ $Type\}$, $P4=\{Color\}$, $P5=\{Number$ $of$ $Doors\}$, $P6=\{Made$ $By\}$, $P7=\{Vehicle$ $Type\}$, $P8=\{Inspect,$ $Mileage,$ $Color\}$, $P9=\{Inspect,$ $Mileage,$ $Color,$ $Number$ $of$ $Doors\}$, $P10=\{Inspect,$ $Mileage,$ $Color,$ $Number$ $of$ $Doors,$ $Number$ $of$ $Seats\}$, $P11=\{Inspect,$ $Mileage,$ $Color,$ $Number$ $of$ $Doors,$ $Number$ $of$ $Seats,$ $Made$ $By\}$.
	\end{itemize}

	\begin{figure}[h!]
		\begin{center}
			\includegraphics[width=0.45\textwidth]{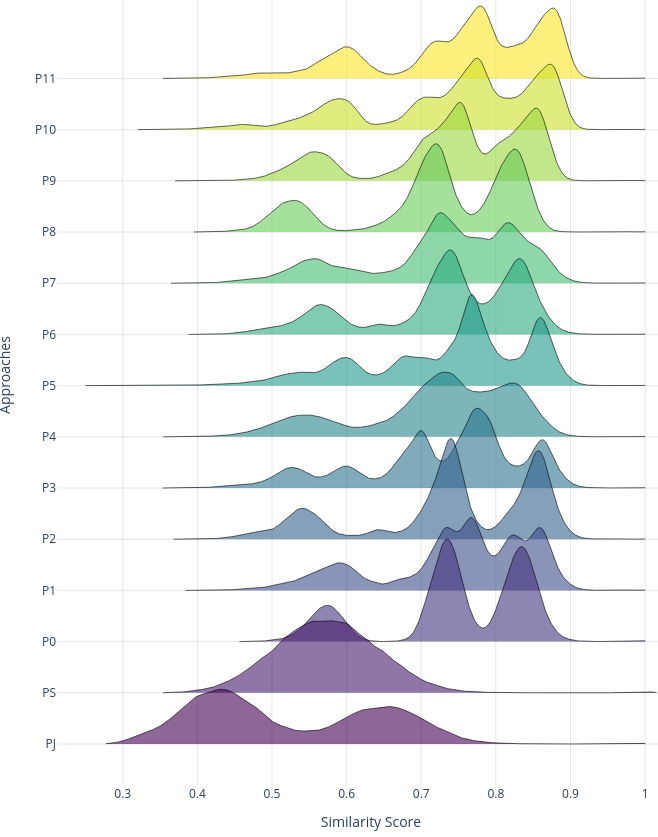}
		\end{center}
		\vspace{-0.5cm}
		\caption{Histogram representing the distribution of similarity scores across different approaches.}
		\label{fig02}
		\vspace{-0.7cm}
	\end{figure}
		
	In our comprehensive study, we evaluated our approach against three other methodologies by calculating similarity scores for each RDF instance of used vehicles against others, using a RDF dataset of 1000 used vehicle instances. Higher similarity scores in measuring RDF graph similarity indicate better experimental results by reflecting a greater correspondence between the structures and relationships in the graphs. When compared to the $PS$, $PJ$, and $P0$ methods, our approach from $P1$ to $P11$ showed notably better results in terms of similarity scores. The visual representation of these results is evident in both a heat map (Fig. \ref{fig03}) and a histogram chart (Fig. \ref{fig02}). The heat map's yellow color distribution indicates higher similarity scores for our approach ($P1$ to $P11$) in comparison to other methods, with a consistent pattern across the map. The histogram chart further confirmed this, as the similarity score distribution for our approach ($P1$ to $P11$) is significantly higher than the others. 
	These findings indicate that our approach, particularly $P11$, achieved the highest maximum similarity score (409765 out of 100000 comparisons), significantly surpassing other methods. Specifically, $P11$ is designed to measure RDF graph similarity by employing weighted properties for various numeric properties of vehicle models. This approach involves assigning weights to properties such as $Inspect$, $Mileage$, $Color$, $Number$ $of$ $Doors$, $Number$ $of$ $Seats$, and $Made$ $By$. The effective use of weighted properties in $P11$ enables a more nuanced and accurate assessment of similarity between vehicle models based on these properties.
	\begin{figure}[h!]
		\begin{center}
			\includegraphics[width=0.45\textwidth]{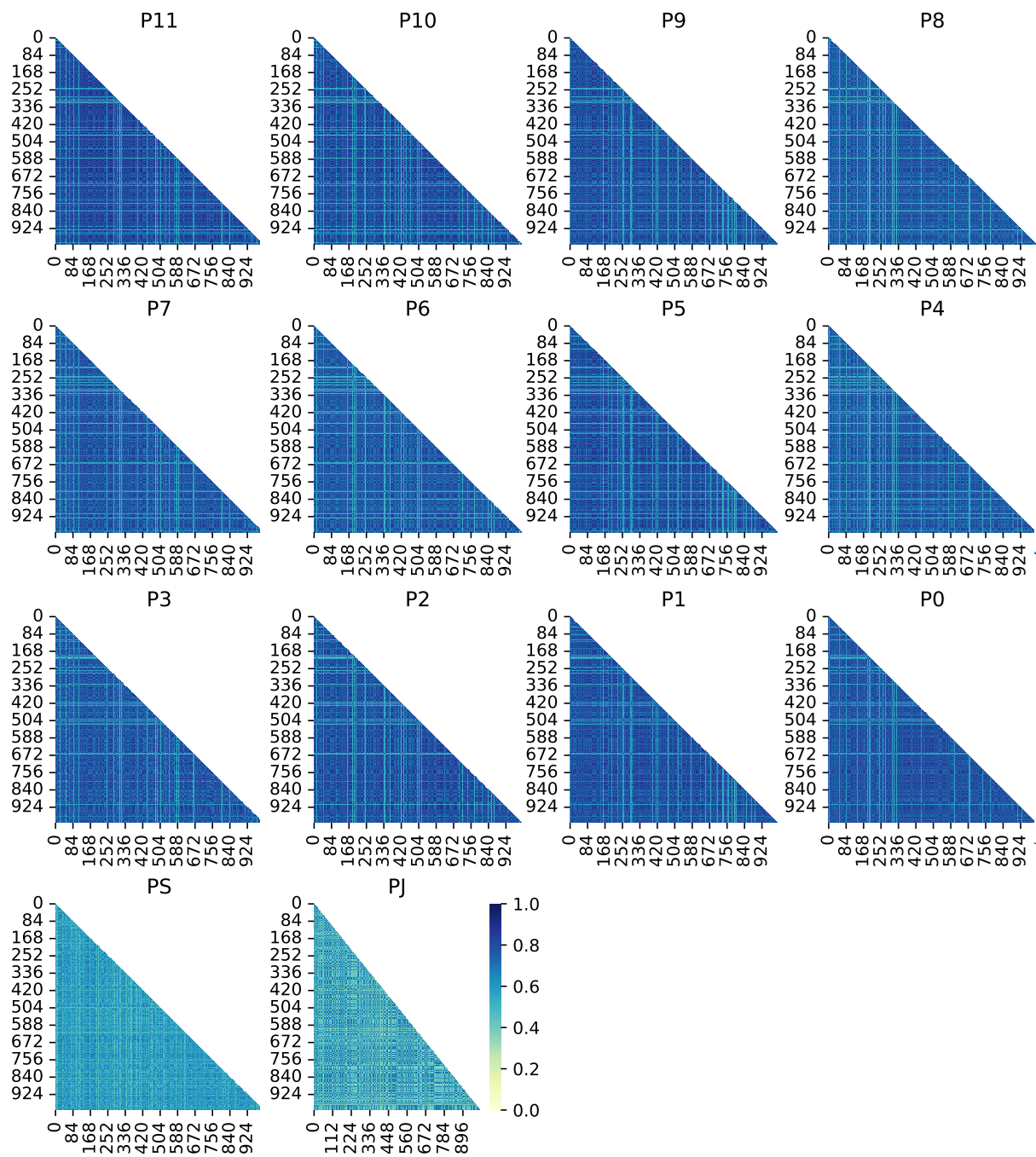}
		\end{center}
		\vspace{-0.5cm}
		\caption{Correlation heat map of similarity measures across 1000 used vehicles for different approaches (\textit{Similarity scores ranging between 0 and 1 imply 0 as entirely distinct and 1 as complete similarity between vehicle items})}
		\label{fig03}
		\vspace{-0.6cm}
	\end{figure}
	
	While the use of weighted properties for measuring similarity has its advantages, as highlighted in the previous section \ref{sec_wp}, our experimental findings point to a number of challenges with this approach. The foremost challenge is the determination of appropriate weights for individual properties, a process often marred by subjectivity. Individuals might assign different weights based on their personal opinions or preferences, leading to variability in results. Additionally, scalability poses a significant challenge. As the quantity of properties and the size of RDF datasets grow, managing and processing these weighted properties becomes increasingly complex and demands more resources. This complexity could render the approach impractical for extremely large datasets. Another critical issue is the inherent subjectivity and potential for bias in weighting. Since the assignment of weights can vary greatly depending on individual judgment, it may lead to inconsistent and potentially biased assessments in measuring similarity.

	\section{Conclusion and Perspectives}
	In this paper, we conducted an investigation to explore weighted property approaches for RDF graph similarity measure. We proposed and validated an approach for measuring RDF graph similarity, emphasizing the role of weighted properties. Through a comprehensive experimental study on an RDF graph dataset in the vehicle domain, our approach demonstrated promising results in improving the accuracy and utility of RDF graph similarity measures, particularly in the automotive industry, where it can significantly enhance the personalization within recommender systems and semantic search tools. We acknowledge the challenges of our work, including determining appropriate weights for each property, addressing subjectivity in the weighting process, and ensuring scalability for large datasets. These challenges have been central to our research. Looking ahead, there is potential for further development and application of this approach in various industries. Extending the approach to other domains and integrating novel and advanced weighted property assignment techniques are promising directions for future research.
	
	
	\section*{Acknowledgment}
	This work was funded by the French Research Agency (ANR) and by the company Vivocaz under the project France Relance - preservation of R\&D employment (ANR-21-PRRD-0072-01).
	
	\vspace*{-0.3cm}
	\bibliographystyle{IEEEtran}
	\bibliography{references} 
	
\end{document}